\documentclass[runningheads]{llncs}
\usepackage[T1]{fontenc}
\usepackage{newtxtext}
\usepackage[varvw]{newtxmath}
\usepackage{graphicx,booktabs}
\usepackage{cite}
\usepackage[hidelinks]{hyperref}
\usepackage[ruled,vlined,linesnumbered]{algorithm2e}

\usepackage[table]{xcolor}
\definecolor{CARhigh}{RGB}{255,255,255}

\hypersetup{pdftitle={AIRPiano: Piano Score Following with Tabletop Hand Gestures}}

\newif\ifanonymous
\anonymousfalse
\title{AIRPiano: Piano Score Following with Tabletop Hand Gestures}
\titlerunning{AIRPiano: Tabletop Piano Score Following}
\ifanonymous
  \author{Anonymous Authors}
  \authorrunning{Anonymous Authors}
  \institute{}
  \hypersetup{pdfauthor={}}
\else
  \author{Jinxuan Zhu \and Zhiyao Duan}
  \authorrunning{J. Zhu and Z. Duan}
  \institute{University of Rochester,\\
    Rochester, USA\\
    \email{jzhu79@u.rochester.edu, zhiyao.duan@rochester.edu}}
  \hypersetup{pdfauthor={Jinxuan Zhu and Zhiyao Duan}}
\fi

\begin{document}
\maketitle

\begin{abstract}
\textup{AIRPiano} aims to bring sound to the common practice of rehearsing piano music on a tabletop. Guided by a known score, users make piano-like finger movements on a flat surface without reproducing absolute keyboard positions. The score supplies pitches, while hand motion determines which score events to play and when. We formulate this interaction as online alignment between tabletop hand motion and symbolic score events. To address the ambiguity of tabletop motion, its differences from piano-playing motion, and the scarcity of aligned motion-score data, we develop motion and score representations, build a data processing pipeline, and train a Transformer with supervised attention on existing piano performance recordings. We further design a streaming inference framework that combines local alignment predictions with sequence tracking for real-time interaction. We construct two evaluation sets, VSF-Piano and VSF-Tabletop, to assess alignment on piano and tabletop performances, respectively. After a post-hoc global timing adjustment, AIRPiano achieves a mean chord alignment rate of 74\% within a 133~ms tolerance on the paired subset of VSF-Tabletop, compared with 79\% on VSF-Piano.

\end{abstract}

\section{Introduction}
\label{sec:introduction}

\begin{figure}[htbp]
    \centering
    \includegraphics[width=\linewidth]{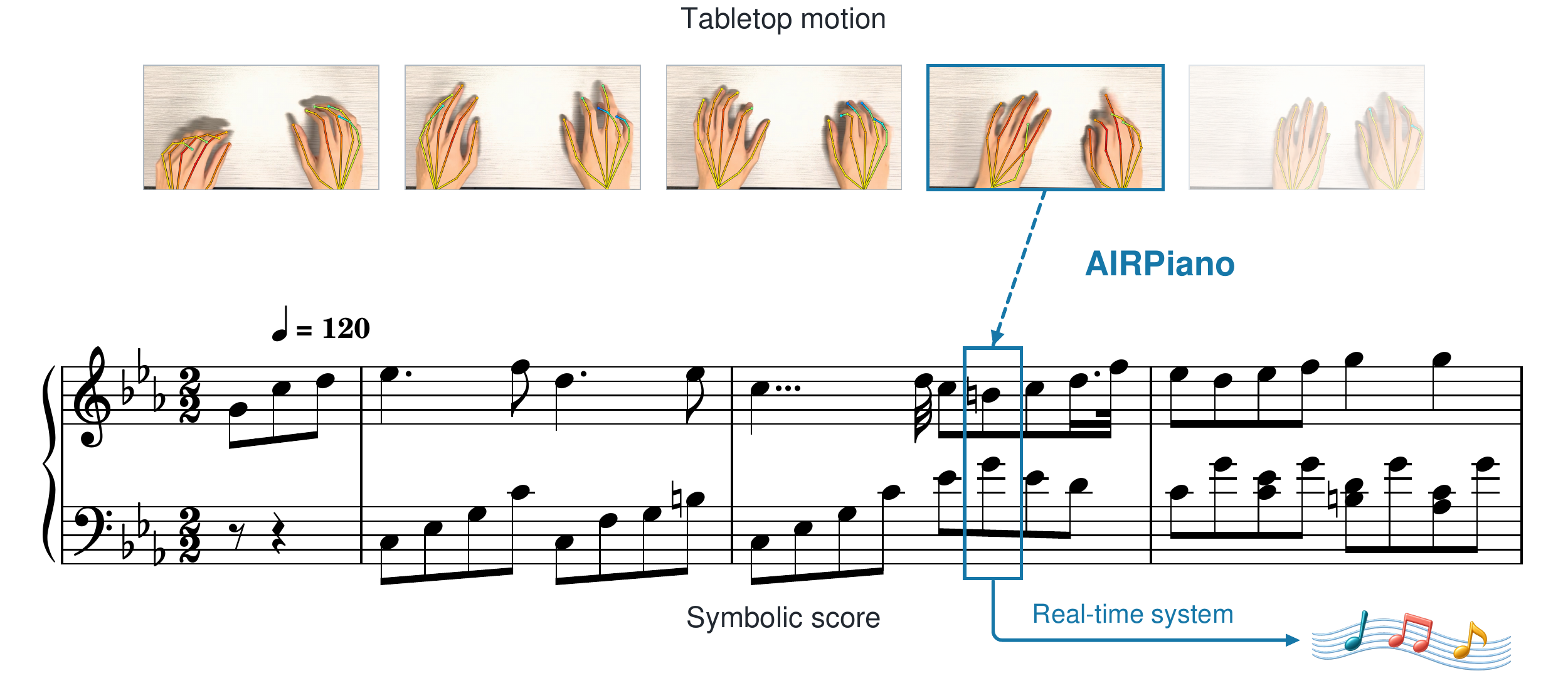}
    \caption{An illustration of our AIRPiano system. The score supplies the pitches, while tabletop hand motion determines which score events to play and when. The dashed arrow links the current video frame to the score event selected for real-time MIDI playback.}
    \label{fig:overview}
\end{figure}

When a piano is unavailable, pianists may still want to rehearse a piece. Under the circumstances, they can tap their fingers on a table or their thighs or even in the air, either while reading a new score or recalling a familiar piece. A system that translates such movements into sound could support practice by helping pianists assess how their approximate finger movements correspond to the score. It could also enable expressive performance without a physical piano or the need to reproduce precise key locations, reducing technical demands particularly in passages involving large leaps.

Many existing virtual piano systems enable music-making without a physical piano by mapping fingertip taps to keys on a virtual or paper keyboard~\cite{yeh2010virtual,liang2016barehanded,vishal2018paperpiano,avola2023keyrtual,venanzi2026musikeyrtual}. However, virtual or paper keyboards may not always be readily available, and a full-sized keyboard layout requires substantial surface space. Without tactile key boundaries, accurately targeting the key corresponding to the intended pitch can also be challenging.

Score-guided systems offer an alternative by supplying pitches from a known score while allowing performers to control aspects of timing and dynamics~\cite{nishimoto2002musical,katayose2004expressive,cancino2019espressione}. Some systems use visual hand tracking and score-aligned rehearsal references to follow performances on a physical keyboard~\cite{ritter2013imuse,pet2024combining}. 

However, directly aligning tabletop hand motion to symbolic score events, without any keyboard or piece-specific rehearsal reference, remains underexplored. We consider the setting where a pianist rehearses a piece from a known score using piano-like finger movements within a limited tabletop area, without aiming for fixed key positions. We formulate this task as online motion-score alignment, a form of visual score following that uses hand motion instead of the more commonly used audio or MIDI inputs.

This alignment task presents several challenges. The main challenge is that tabletop movements need not preserve the absolute spatial arrangement of pitches on a full keyboard. Tap positions may follow the keyboard's relative pitch ordering within a passage, but the same tabletop area can be reused across different pitch ranges, making absolute hand position an unreliable cue for alignment. Furthermore, hand motion and symbolic scores differ substantially in their representations and temporal structure. Paired motion-score data are scarce, and motion in existing piano performance recordings differs from that in tabletop interaction. Addressing these challenges could enable immediate musical feedback during tabletop practice and provide a foundation for richer expressive control, such as per-note velocity control.

In this paper, we introduce \textup{AIRPiano} to address these challenges.\footnote{A demonstration is available at \url{https://anonymous.4open.science/w/AIRPiano-90AC/}. The code and full dataset will be released upon acceptance.} Our main contributions are:
\begin{itemize}
    \item We formulate score-guided audible tabletop performance as an online motion-score alignment task and construct \textup{VSF-Piano} and \textup{VSF-Tabletop} evaluation sets for this new task with an event-level evaluation protocol.
    \item We build a data processing pipeline to extract aligned motion and score data from existing piano performance recordings, develop representations that facilitate transfer from piano performance to tabletop interaction, and train a Transformer with supervised attention to learn correspondences between the two modalities.
    \item We design a streaming inference framework that combines local alignment predictions with sequence tracking for real-time interaction.
\end{itemize}
\section{Related Work}
\label{sec:related_work}

\subsection{Score-Guided Visual Pianos}
\subsubsection{Vision-based virtual pianos.}
Many studies have explored virtual performance interfaces for different instruments. The piano is a popular example, and advances in computer vision have made it possible to capture piano-like gestures using widely available cameras or infrared hand-tracking devices. Some systems detect virtual keystrokes by tracking fingertip movements in single-view video or by detecting taps on an unmarked flat surface with an RGB-D camera~\cite{yeh2010virtual,liang2016barehanded}. Others use a keyboard drawn on paper to inform players which tap locations correspond to which pitches~\cite{vishal2018paperpiano,avola2023keyrtual,venanzi2026musikeyrtual}.

On a physical piano, a camera can locate individual keys and detect when they are pressed~\cite{kang2019virtual}. Visual piano transcription systems can therefore estimate the pitches played on a real piano from video alone~\cite{su2020audeo,kim2026v2n}. A flat surface, however, provides no tactile boundaries between keys and no visible change in key state. This makes it harder for the player to specify a pitch and for the system to identify it. Matsusue et al.\ try to address this problem by combining fingertip position with a given chord progression to correct the detected pitch, but the choice is still limited to the key at the detected tap position or a neighboring key~\cite{matsusue2017touch}. Piano Genie takes a different approach for improvisation, using a learned decoder to map presses on eight buttons to piano notes~\cite{donahue2019pianogenie}. These approaches do not use a full score to guide performance.

\subsubsection{Score-guided performance interfaces.}
Conducting is the art of guiding an ensemble’s interpretation of a musical score. Virtual conducting gives users a similar role, letting them shape tempo or dynamics of a performance given a score or a recording. \emph{Personal Orchestra} and \emph{You're the Conductor} track baton movements to control the playback of orchestral recordings~\cite{borchers2004personal,lee2004conductor}. \emph{Home Conducting} maps movements captured by a webcam to expressive changes in the performance of a score~\cite{friberg2005home}. Mermerci et al.\ estimate the conductor's phase within a bar from body motion using an LSTM, then use the resulting beat cues to adjust playback speed~\cite{mermerci2026realtime}.

The same idea of letting a performer shape a known piece also appears in score-guided piano interfaces. In \emph{Coloring-in Piano}, each keypress triggers the next pitch in a stored sequence~\cite{nishimoto2002musical}. iFP combines the performer's beat timing and dynamics with a performance template that shapes the notes within each beat~\cite{katayose2004expressive}. In \emph{Con Espressione!}, hand movements control overall tempo and loudness, while a model predicts local variations in timing, dynamics, and articulation for each note in the score~\cite{cancino2019espressione}.

The piano gesture-tracking component of IMuSE is closer to our setting because it follows the hand movements of a pianist. To synchronize the accompaniment, it matches live hand movements to prerecorded rehearsal trajectories aligned with the score~\cite{ritter2013imuse}. However, this requires a reference trajectory for each piece rather than the symbolic score alone, and the pianist performs on a physical keyboard rather than a tabletop.

\subsection{Score Following}
Performance-score alignment links events in a score to the times at which they are played. Offline methods can use the entire performance to match individual notes, as in TheGlueNote, which learns note similarities between MIDI sequences~\cite{peter2024gluenote}. Online score following tracks the musician as they play, enabling applications such as automatic accompaniment. For symbolic input, approaches range from Dannenberg's incremental matching of incoming notes~\cite{dannenberg1984online} to a tracking policy trained with offline reinforcement learning~\cite{peter2023online}. Audio-based systems can track score position online from acoustic
features~\cite{park2025matchmaker} or first transcribe the performance and then align the resulting note events to the score~\cite{peter2025pairing}.

Compared with audio-based score following, video-based score following has received less attention. Nevertheless, video can provide evidence of score position when the audio is ambiguous or absent. Like IMuSE~\cite{ritter2013imuse}, the system of Pet and Maezawa uses a score-aligned rehearsal reference. It combines audio with hand and fingertip positions along the keyboard to improve tracking in noise, and was also evaluated using hand landmarks alone~\cite{pet2024combining}. Movement can also provide an early cue to timing. As future work, Raphael and Gu proposed using piano-key motion to anticipate note strikes when the piano and orchestra enter together~\cite{raphael2009orchestral}. Research also shows that pianists exhibit pre-contact finger motion during tabletop tapping~\cite{loehr2007tapping}. These movements could give an interactive system time to process a gesture before the intended onset.

However, directly aligning tabletop hand motion with a symbolic piano score remains underexplored. On a tabletop, the same tap location can correspond to different pitches, introducing spatial ambiguity that existing audio-based and keyboard-based visual score-following methods do not address.
\section{Method}
\label{sec:method}

AIRPiano uses a Transformer to combine observed hand motion with pitch and rhythm patterns in the score and predict local motion-score alignments. Its streaming inference framework combines these predictions with sequence tracking to follow the pianist through the full score and schedule chord playback. The system predicts upcoming chord onsets from the observed motion history to allow for inference latency.

\begin{figure}[htbp]
    \centering
    \includegraphics[width=\linewidth]{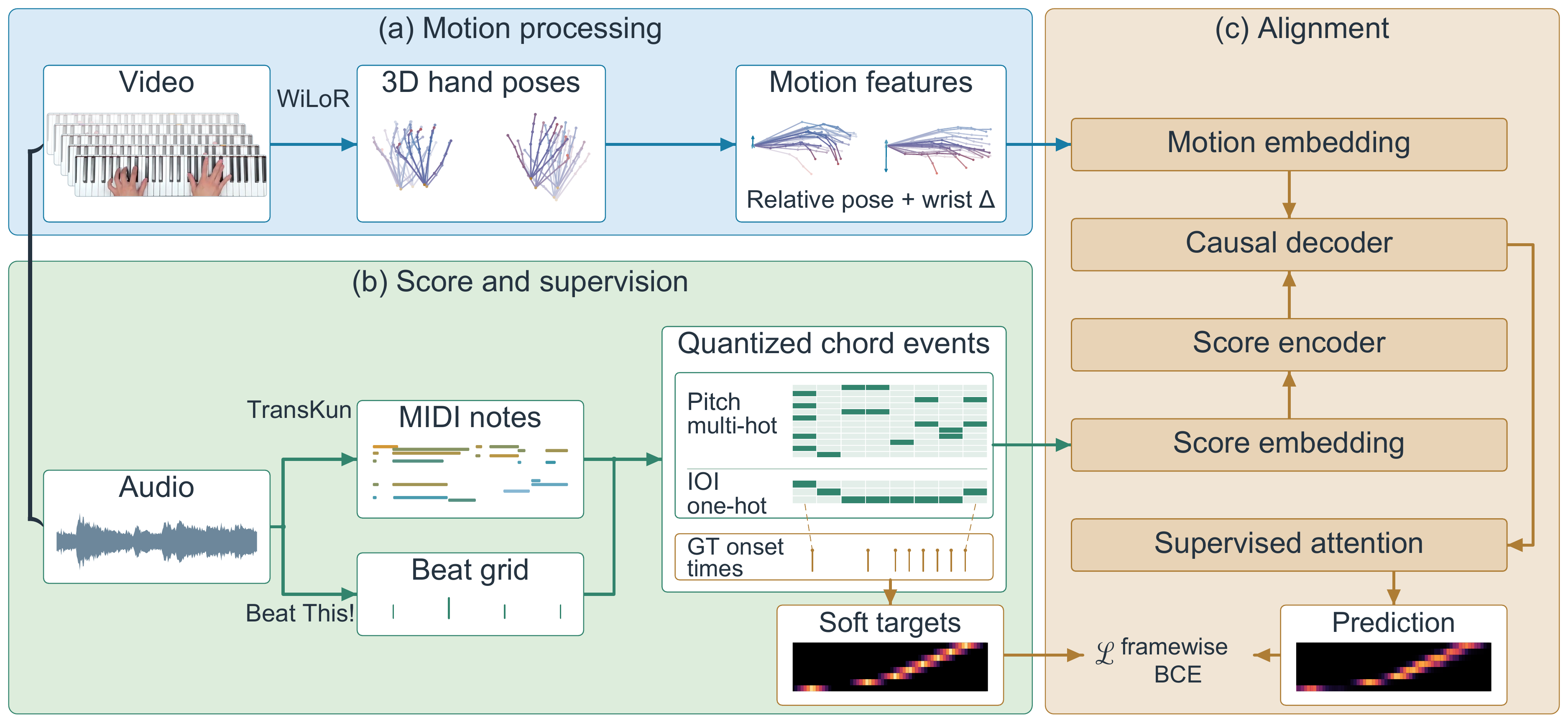}
    \caption{Overview of AIRPiano's data processing and training framework. (a) WiLoR extracts 3D hand poses from video, which are converted into wrist-relative joint positions and wrist displacements. (b) TransKun~\cite{yan2024scoring} and \emph{Beat This!}~\cite{foscarin2024beat} extract MIDI notes and a beat grid from the corresponding audio. MIDI notes are quantized and grouped into chord-onset events, represented by pitch multi-hot and IOI one-hot vectors for model input, while original performance onset times provide soft alignment targets. (c) A score encoder and a causal motion decoder produce frame-to-chord alignment predictions through supervised attention. The model is trained using framewise binary cross-entropy between the predictions and soft targets.}
    \label{fig:learning_framework}
\end{figure}

\subsection{Problem Formulation}
\label{sec:problem}

As discussed above, we formulate the task of AIRPiano as online motion-score alignment. We currently focus on onset events because they mark when notes begin and are the basic timing events needed to trigger score-guided playback. Let $S=(s_1,\ldots,s_L)$ denote a known sequence of $L$ chord-onset events. Notes with the same quantized onset position in the score are grouped into chords, and events are ordered by these positions.

Let $m=(m_0,\ldots,m_{N-1})$ denote a sequence of $N$ hand motion video frames received sequentially. We represent the alignment at frame $t$ by a set $A_t$ containing the events assigned to that frame,
\begin{equation}
    A_t\subseteq\{s_1,\ldots,s_L\}.
\end{equation}
At each frame, zero, one, or multiple events may be triggered. An ideal alignment satisfies uniqueness, coverage, and monotonicity: specifically, $A_u\cap A_v=\varnothing$ for $u\neq v$, $\bigcup_t A_t=\{s_1,\ldots,s_L\}$, and $s_i\in A_u$ and $s_j\in A_v$ with $i<j$ imply $u\leq v$. Nevertheless, the model is not explicitly constrained to satisfy all these properties.

At streaming step $t$, the system uses the observed video history $m_{\leq t}=(m_0,\ldots,m_t)$ and the full score $S$ to predict which events should be triggered $\delta$ frames later, i.e., the set $A_{t+\delta}$. Here $\delta\in\mathbb{Z}_{\geq 0}$ denotes the anticipation in frames, following~\cite{pasini2026liveband}. At a frame rate of $f$, which we currently fix at 30 frames per second, the corresponding anticipation is $\tau=\delta/f$ seconds. A positive $\tau$ provides time for motion processing, model inference, and sound preparation before the predicted onset.

\subsection{Motion and Score Representations}
\label{sec:representations}
\subsubsection{Motion.}
Instead of feeding a video frame $m_t$ directly into the model, we first extract hand poses to remove keyboard and background appearance from the input and provide a compact representation for the alignment backbone. We use WiLoR~\cite{potamias2024wilor}, a recent state-of-the-art hand localization and reconstruction model, to estimate 3D poses for both hands, each represented by 21 joints, following the tradition of MediaPipe~\cite{zhang2020mediapipe}. For each detected hand, WiLoR outputs a complete set of joint estimates in physical units, inferring invisible joints when the hand is only partially visible, and marks a hand as missing only when it is not detected at all.

Let $\mathbf{x}_{t,h,j}\in\mathbb{R}^3$ denote the estimated position of joint $j$ of hand $h$ at frame $t$ in metric camera coordinates, expressed in decimeters, where $h\in\{\mathrm{left},\mathrm{right}\}$ and $j=0$ denotes the wrist. We also define a binary presence flag $b_{t,h}\in\{0,1\}$, where $b_{t,h}=1$ iff hand $h$ is detected at frame $t$, regardless of whether all its joints are visible. When $b_{t,h}=0$, we set $\mathbf{x}_{t,h,j}=\mathbf{0}$ for all joints. For the 20 non-wrist joints, we compute their positions relative to the wrist,
\begin{equation}
    \widetilde{\mathbf{x}}_{t,h,j}=\mathbf{x}_{t,h,j}-\mathbf{x}_{t,h,0}, \qquad j=1,\ldots,20.
\end{equation}
Since similar finger movements can occur at different tabletop locations, this representation captures local articulation while reducing reliance on absolute hand position.

To retain information about hand translation, such as movement during large leaps, we also compute the wrist displacement between consecutive frames,
\begin{equation}
    \Delta\mathbf{x}_{t,h,0}=
    \begin{cases}
        \mathbf{x}_{t,h,0}-\mathbf{x}_{t-1,h,0}, & \text{if } t>0,\ b_{t,h}=b_{t-1,h}=1,\\
        \mathbf{0}, & \text{otherwise}.
    \end{cases}
\end{equation}

Each hand is therefore represented by wrist displacement in joint slot 0 and wrist-relative positions in the remaining 20 slots, as well as the presence flag $b_{t,h}$.

\subsubsection{Score.}
For each chord-onset event $s_i$, its pitch content is encoded as a 128-dimensional multi-hot vector $\mathbf{p}_i$, whose entries indicate the presence of MIDI pitches from 0 to 127. To represent the score's rhythmic structure, we encode the interval from each event to the next event in score time a.k.a. beat, which allows the model to use relative timing cues alongside pitch content when aligning hand movements with the score. Let $q_i$ denote the quantized onset tick of event $s_i$, measured at 24 ticks per beat to support common binary and triplet subdivisions without rounding. Its inter-onset interval (IOI) feature is
\begin{equation}
    d_i=q_{i+1}-q_i,\qquad i=1,\ldots,L-1.
\end{equation}
For the final event, the interval extends to the latest offset of its notes. We clamp $d_i$ to $[1,96]$ ticks and encode it as a 96-dimensional one-hot vector $\mathbf{r}_i$.

\subsection{Cross-Modal Alignment Learning}
\label{sec:alignment_learning}

\subsubsection{Model.}
For each frame $m_t$, we process it as described above, flatten $\Delta\mathbf{x}_{t,h,0}$ and $\widetilde{\mathbf{x}}_{t,h,j}$ for both hands, and append $b_{t,h}$, yielding a 128-dimensional vector. An MLP embedding block projects it into a 512-dimensional token containing information from both hands. For each event $s_i$, the pitch vector $\mathbf{p}_i$ and IOI vector $\mathbf{r}_i$ are separately projected into 512-dimensional embeddings and added together. Each embedding block consists of a linear layer followed by GELU, layer normalization, and dropout.

We use a Transformer encoder-decoder architecture~\cite{vaswani2017attention}, with sinusoidal positional encodings added to the motion and score tokens. The score tokens pass through four encoder layers with bidirectional self-attention; the motion tokens pass through four decoder layers with causal self-attention over motion frames and cross-attention to the encoded score. Thus, each motion token incorporates the observed motion history and the candidate score context.

Motivated by attention-based alignment in machine translation~\cite{garg2019jointly} and speech synthesis~\cite{ren2019fastspeech}, we use a single supervised attention-style scoring head for motion-score alignment. For an input pair containing $T$ frames and $C$ candidate events, let $\mathbf{u}_t$ and $\mathbf{v}_i$ denote the resulting motion and score representations. The head computes an alignment matrix $\widehat{\mathbf{P}}\in[0,1]^{T\times C}$ using attention-style scaled dot products,
\begin{equation}
    \widehat{P}_{t,i}=\sigma\left(\frac{g_m(\mathbf{u}_t)^\top g_s(\mathbf{v}_i)}{\sqrt{d_a}}+\beta\right),
\end{equation}
where $g_m$ and $g_s$ are learned query and key projections to $d_a=32$ dimensions, $\beta$ is a learned scalar bias, and $\sigma$ is the sigmoid function. We use an entrywise sigmoid instead of softmax normalization to allow zero, one, or multiple events to receive high scores at each motion frame. We still refer to this output as the supervised attention matrix.

\subsubsection{Supervision.}
\begin{figure}[t]
    \centering
    \includegraphics[width=0.8\linewidth]{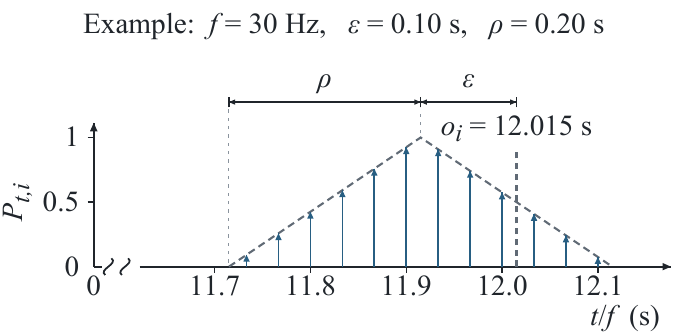}
    \caption{An example of triangular onset supervision. The smooth profile reduces abrupt label changes under small timing shifts, while a positive \(\varepsilon\) encourages the model to use preparatory hand movements for anticipatory alignment. Both can support the prediction of \(A_{t+\delta}\) from the observed motion history \(m_{\leq t}\).}
    \label{fig:onset_targets}
\end{figure}

We construct a soft target matrix $\mathbf{P}\in[0,1]^{T\times C}$ from aligned performances. Following the onset regression targets of Kong et al.~\cite{kong2021highresolution}, we use triangular temporal targets. Let $o_i$ denote the reference onset time of event $s_i$ in seconds. Using the global motion-frame index $t$ and frame rate $f$, we define
\begin{equation}
\label{eq:onset_target}
    P_{t,i}=\max\left(0,\,1-\frac{\left|t/f-(o_i-\varepsilon)\right|}{\rho}\right),
\end{equation}
where $\rho>0$ is the temporal radius and $\varepsilon$ is a temporal offset. Figure~\ref{fig:onset_targets} illustrates the resulting framewise targets and the motivation behind this design.

We supervise $\widehat{\mathbf{P}}$ with $\mathbf{P}$ using binary cross-entropy averaged over valid frame-event pairs, excluding padded frames and events. Each predicted entry $\widehat{P}_{t,i}$ is therefore an onset score ranging from 0 to 1, trained to follow the temporal target for event $s_i$ at frame $t$. During streaming inference, these continuous scores are converted into discrete decisions about which score events to trigger.

\subsection{Streaming Inference}
\label{sec:streaming}

As described above, the model is trained on local score windows and their corresponding motion segments. To apply it to streaming hand motion of arbitrary duration, we design a streaming inference algorithm, inspired by Dannenberg's incremental matching and sliding window strategy~\cite{dannenberg1984online}. The algorithm maintains a score window $S_{\mathrm{window}}$, a motion history $M$ with starting frame index $t_{\mathrm{begin}}$, and an anchor-time map $\mathcal{A}$. Upon receiving each frame $m_t$, it runs the model on the score and motion windows, triggers events from the onset scores at frame $t$, and schedules MIDI playback at the predicted onset times. It then uses DP tracking to decide whether to accept each triggered event as a timing anchor and updates $S_{\mathrm{window}}$ accordingly. After each window update, $t_{\mathrm{begin}}$ is updated to match the beginning of $S_{\mathrm{window}}$ using available anchors in $\mathcal{A}$. This process continues until the input motion frame stream ends. The main operations are detailed below.

\begin{itemize}
    \item \textbf{Triggering of events.} At frame $t$, the last row of the predicted alignment matrix $\widehat{\mathbf{P}}$ provides the onset scores for events in $S_{\mathrm{window}}$. An event $s_i$ is triggered when its score exceeds $0.5$, and triggering is re-enabled only after the score falls below $0.5$. This prevents a sustained response from generating repeated triggers. On the rising branch of the triangular profile defined in Eq.~\eqref{eq:onset_target} and illustrated in Fig.~\ref{fig:onset_targets}, solving for a target value of $0.5$ gives the crossing time $o_i-\varepsilon-\rho/2$, which is $(\varepsilon+\rho/2)$ seconds ahead of $o_i$. Ideally, a nominal anticipation of $\tau=\delta/f$ seconds can therefore be obtained by choosing
\begin{equation}
    \varepsilon+\frac{\rho}{2}=\frac{\delta}{f}=\tau.
\end{equation}
The system then schedules the pitches of each triggered event $s_i$ for playback at $\widehat{o}_i=(t+\delta)/f$ with a fixed velocity and duration. Due to processing overhead, the scheduling decision for frame $t$ is made after its timestamp $t/f$, but ideally before the scheduled onset $\widehat{o}_i$, leaving time to prepare playback.

\item \textbf{Dynamic programming.} Following Dannenberg's adaptation of the longest common subsequence (LCS) problem for online score following~\cite{dannenberg1984online}, we use dynamic programming (DP) to incrementally match triggered event indices against the sequence of score event indices $(1,2,\ldots,L)$. Specifically, for a triggered event index $i$, let $D_j$ and $D'_j$ denote the DP scores at score position $j$ before and after the update, with initial scores $D_j=0$ and boundary value $D'_0=0$. We update, for $j=1,\ldots,L$,
\begin{equation}
    D'_j=
    \begin{cases}
        \max\{D_j,\ D'_{j-1}-1,\ D_{j-1}+1\}, & j=i,\\
        \max\{D_j,\ D'_{j-1}-1\}, & j\neq i.
    \end{cases}
\end{equation}
Matching an event contributes $1$, skipping an event costs $1$, and an extra triggered event can be ignored without penalty. We accept a triggered event only when the DP update produces a new historical best. Following Dannenberg's window-update strategy, an accepted event supplies a timing anchor $\mathcal{A}[i]=\widehat{o}_i$ and recenters $S_{\mathrm{window}}$ around $s_i$; otherwise, the trigger still advances the window by one score event.

\item \textbf{Motion-history start estimation.} To remain consistent with the training setup, we need to align the beginnings of the motion context and score window after each window update. To be more specific, we estimate the onset $\widetilde{o}_\ell$ of the first event $s_\ell=S_{\mathrm{window}}[1]$. We use $\mathcal{A}[\ell]$ if available; otherwise, we apply linear interpolation between surrounding anchors or extrapolation from the nearest endpoint pair using the target position $q_\ell$ and the anchors' score positions. The starting frame index is hence updated as $t_{\mathrm{begin}}=\lfloor f\widetilde{o}_\ell\rfloor$, clamped to $[0,t]$.
\end{itemize}

\section{Training}
\label{sec:training}

\subsection{Training Data}
\label{sec:training_data}

As discussed above, to our knowledge, no public tabletop motion-score dataset is available. Piano performance datasets with symbolic scores are also scarce, with recent examples such as SKY-Piano~\cite{bae2026skypiano} containing 11 hours of recordings. Synchronized piano video and audio, such as top-view recordings of piano performances, are more readily available. We use PianoMotion10M~\cite{gan2025pianomotion10m}, which contains 116 hours of top-view piano performance videos collected from Bilibili. We observe that these recordings predominantly feature popular music and supplement them with classical piano performances from PianoYT~\cite{koepke2020sight}, a dataset containing over 20 hours of piano performance videos collected from YouTube. Using the provided links, we download videos together with their corresponding audio. We select videos at a resolution of $1920\times1080$ pixels and 30 frames per second where available, resampling when necessary and excluding inaccessible recordings.

We use the training split of PianoYT and both the training and test splits of PianoMotion10M for training. Since we have our own test set, this allows us to make fuller use of the available recordings, particularly the substantial test split of PianoMotion10M, which is even larger than its training split. We retain the validation split of PianoMotion10M for in-domain evaluation.

\subsection{Data Preparation}
\label{sec:data_preparation}

As illustrated in Fig.~\ref{fig:learning_framework}, motion processing follows Section~\ref{sec:representations}, using the same pipeline during training and inference. Meanwhile, the score representation in Section~\ref{sec:representations} assumes chord-onset events with pitch content and quantized score positions, which are not directly available in datasets with only synchronized piano video and audio. We therefore try to transcribe approximate scores by quantizing transcribed performance MIDI against an estimated beat grid. We also retain the original performance onset times to derive the reference onsets $o_i$ required for alignment supervision in Section~\ref{sec:alignment_learning}.

Specifically, we transcribe each recording's audio into MIDI notes with onset and offset times using TransKun 2.0~\cite{yan2024scoring} and estimate beat times using \emph{Beat This!}~\cite{foscarin2024beat} with DBN postprocessing (ignoring downbeat labels) as shown in Fig.~\ref{fig:learning_framework}(b). We use 24 ticks per beat, restricting candidate offsets within each beat to $\mathcal{R}=\{0,\allowbreak 3,\allowbreak 4,\allowbreak 6,\allowbreak 8,\allowbreak 9,\allowbreak 12,\allowbreak 15,\allowbreak 16,\allowbreak 18,\allowbreak 20,\allowbreak 21\}$ to cover common binary and triplet subdivisions. Candidate timestamps are obtained by linear interpolation between detected beats, and each transcribed note onset and offset is mapped to the candidate tick with the nearest timestamp. Notes sharing the same quantized onset form an event $s_i$ at tick $q_i$. Its reference onset $o_i$ is the mean of its notes' original transcribed onset times.

\subsection{Sampling and Augmentation}
\label{sec:sampling}

We sample overlapping score windows of $C=50$ chord events with a stride of 12 events and extract the corresponding motion segments using their reference onset times. We discard windows in which neither hand is detected in more than $20\%$ of the frames, helping filter out invalid segments such as video intros and outros.

For motion augmentation, we independently perturb the start and end of each motion segment by up to $0.5$ seconds to reduce dependence on precise window boundaries. For score augmentation, we scale all IOI values within each segment by a shared factor sampled uniformly from $\{0.5, 1, 2\}$ to reduce reliance on their absolute scale.

\subsection{Model Configuration and Training Setup}
\label{sec:training_setup}

Each Transformer layer uses eight attention heads and a feed-forward dimension of 2048, with GELU activation, pre-layer normalization, and dropout of $0.1$. For the triangular supervision in Eq.~\eqref{eq:onset_target}, we set $\rho=0.2$ seconds and $\varepsilon=0.1$ seconds, corresponding to a nominal anticipation of $\tau=\varepsilon+\rho/2=0.2$ seconds.

We train for 20 epochs using AdamW with a peak learning rate of $3\times10^{-4}$, weight decay of $0.01$, and gradient norm clipping at $1.0$. We use linear warmup over the first $5\%$ of optimizer steps followed by cosine decay. Training uses BF16 mixed precision and an effective batch size of 16, taking approximately 8 hours on a single NVIDIA RTX 4090 GPU with 24 GB of memory. We use the final-epoch checkpoint for evaluation.
\section{Evaluation}
\label{sec:evaluation}

\subsection{VSF-Piano and VSF-Tabletop}
\label{sec:vsf_datasets}

\begin{figure}[t]
    \centering
    \includegraphics[width=\linewidth]{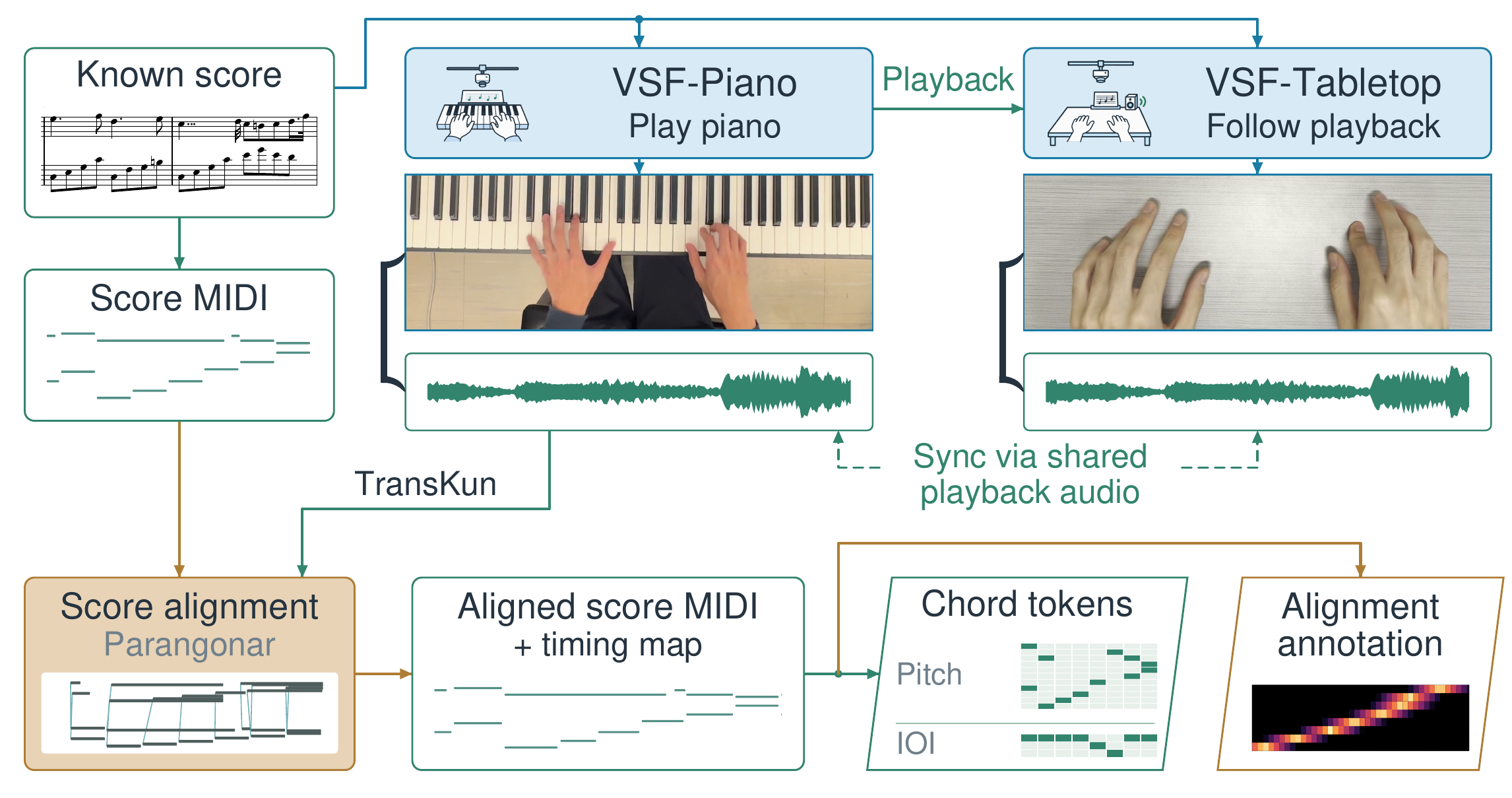}
    \caption{Collection and annotation of VSF-Piano and VSF-Tabletop. Given a score of a known piece, a piano performance is first recorded; each reference piano performance is followed by two tabletop participants, including the performer, who read the same score and try to match the timing of the playback using natural piano-like finger movements on the tabletop, without needing to reproduce absolute keyboard positions. The recordings are synchronized using the shared playback audio to transfer reference onset times from piano to tabletop recordings. We then use TransKun~\cite{yan2024scoring} to transcribe the piano audio and Parangonar~\cite{peter2023online} to align the score MIDI with the transcription, allowing for minor performance or transcription errors. The resulting aligned score MIDI contains score notes at estimated performance times for alignment annotation, while their quantized positions in the score are preserved for chord tokens through a timing map. (During real-time inference, quantized chord tokens can be obtained directly from score MIDI with marked beats; here, the aligned score MIDI also provides reference onset times for evaluation.)}
    \label{fig:testset_pipeline}
\end{figure}

Ground-truth motion-score alignments are difficult to obtain for tabletop performances, which produce no pitched sound. We therefore design a recording and annotation protocol in which participants follow playback of reference piano performances, allowing reference onset times to serve as pseudo ground truth. Using this protocol, illustrated in detail in Fig.~\ref{fig:testset_pipeline}, we construct two test sets, VSF-Piano and VSF-Tabletop, where VSF stands for visual score following. VSF-Piano contains piano recordings from two pianists, while VSF-Tabletop contains tabletop recordings from four participants, including these two pianists. All four participants are amateur pianists who can read piano scores, and the pieces were selected to be within their piano playing abilities.

We select 10 pieces covering different musical styles, tempi, rhythmic patterns, and musical textures. Ornaments such as trills are explicitly realized as note sequences in the score MIDI. Since these well-known pieces may also appear in the source training corpora, we compare MIDI sequences using flexible dynamic time warping (DTW)~\cite{bukey2023flexdtw}, accounting for tempo differences and transposition, and exclude the identified overlapping recordings from training. Recording durations range from approximately 30 to 150~s, covering both short and extended performances. All test videos are recorded from an overhead view at 1080p and 30~fps, with the line joining the wrists approximately parallel to the bottom edge of the video frame, similar to the training data. Backgrounds and exact camera viewpoints may vary across participants.

\subsection{Other Evaluation Set}
As mentioned above, we also use the held-out validation split of PianoMotion10M for in-domain alignment evaluation on piano-playing recordings from the same source as the training data. We construct 30, 60, and 120~s variants, denoted PM30, PM60, and PM120, respectively, to assess alignment performance over different sequence lengths. For each duration, we randomly extract one clip from each sufficiently long recording. Model inputs and reference onset times are obtained using the same data processing pipeline as for training.

\subsection{Evaluation Metric}
\label{sec:metrics}
It is challenging to establish a directly comparable baseline for our tabletop motion-score alignment setting. Nevertheless, we adapt alignment evaluation from conventional score following~\cite{cont2007evaluation,duan2011state} to provide an interpretable measure of absolute alignment performance. Following the alignment-rate definition, we use a chord-level variant, termed chord alignment rate (C-AR), as our main evaluation metric, defined as the proportion of reference chord events aligned within a specified timing tolerance. Specifically, given reference alignment sets $A_t$ obtained by quantizing onset times $o_i$ to the frame grid and predicted alignment sets $\widehat{A}_t$, let $t_i$ and $\widehat{t}_i$ denote the reference and predicted onset frames of event $s_i$, i.e., $s_i\in A_{t_i}$ and $s_i\in\widehat{A}_{\widehat{t}_i}$. For a tolerance of $k$ frames,
\begin{equation}
    \mathrm{C\text{-}AR}@k=\frac{1}{L}\sum_{i=1}^{L}\mathbf{1}\!\left[\left|\widehat{t}_i-t_i\right|\leq k\right],
\end{equation}
where $\mathbf{1}[\cdot]$ is the indicator function and $L$ is the number of reference chord-onset events. If an event has no predicted alignment, we set $\widehat{t}_i=+\infty$ so that it counts as a failure; if multiple predicted alignments exist, we use the earliest frame. We report the mean C-AR across recordings (macro-averaging), giving each performance equal weight.
\section{Results}
\label{sec:results}

\subsection{Main Results}
\label{sec:main_results}

\begin{table}[t]
    \centering
    \caption{Macro-averaged chord alignment rates (C-AR, \%) on VSF-Piano, VSF-Tabletop, VSF-Tabletop (paired), PM30, PM60, and PM120, with tolerances of 2, 4, 8, and 16 frames. VSF-Tabletop (paired) includes tabletop recordings from the same two pianists as VSF-Piano, reducing differences in participant composition. For rows marked $+5$ frames, reference onset times are shifted later by five frames (approximately 166.67~ms at 30 fps), with predictions unchanged.}
    \label{tab:main_results}
    \small
    \setlength{\tabcolsep}{4pt}
    \begin{tabular}{lrrrr}
    \toprule
    Evaluation set & C-AR@2 & C-AR@4 & C-AR@8 & C-AR@16 \\
    \midrule
    VSF-Piano & 6.10 & 26.64 & 80.71 & 88.90 \\
    VSF-Tabletop & 9.24 & 26.71 & 67.77 & 76.17 \\
    VSF-Tabletop (paired) & 6.20 & 23.20 & 73.08 & 83.82 \\
    PM30 & 64.25 & 88.20 & 96.06 & 96.96 \\
    PM60 & 63.48 & 87.29 & 96.09 & 96.92 \\
    PM120 & 65.57 & 89.01 & 96.83 & 97.37 \\
    \midrule
    VSF-Piano (+5 frames) & 58.83 & 79.46 & 85.98 & 88.79 \\
    VSF-Tabletop (+5 frames) & 44.40 & 63.94 & 72.83 & 76.06 \\
    VSF-Tabletop (paired, +5 frames) & 52.05 & 74.03 & 81.26 & 83.87 \\
    \bottomrule
\end{tabular}
\end{table}

As shown in Table~\ref{tab:main_results}, our model achieves C-AR@8 above 96\% on the in-domain evaluation sets PM30, PM60, and PM120. Performance remains similar across these durations, suggesting stable streaming alignment over the evaluated sequence lengths. C-AR@8 decreases to 80.71\% on VSF-Piano, suggesting a domain gap between the source recordings and our collected performances. On VSF-Tabletop, C-AR@8 is 67.77\%, while the paired subset achieves 73.08\%, with a relative decrease of only 9.45\% compared with VSF-Piano, supporting the model's zero-shot transfer from piano to tabletop motion.

C-AR@2 and C-AR@4 are substantially lower on the VSF datasets than on the PM datasets. We search for the integer-frame shift of reference onset times that maximizes macro-averaged C-AR@4. A shift of $+5$ frames, moving the reference onsets later, is optimal for both VSF-Piano and VSF-Tabletop. This shift markedly increases C-AR@2 and C-AR@4, suggesting a systematic timing offset, with predictions tending to occur approximately five frames after the reference onsets. No comparable offset is observed on PM60, suggesting that the bias is likely related to differences between the training and VSF data. Differences in audio delays relative to video across recording devices may contribute to this bias, as reference onset times are derived from audio rather than video, and actual gesture onsets are not independently measured. Due to these uncertainties, which may particularly affect evaluation at strict tolerances, we use C-AR@8 with the original, unshifted reference onset times for the main analysis.

\subsection{Analysis across Participants and Pieces}
\label{sec:participant_piece_analysis}
\begin{figure}[t]
    \centering
    \includegraphics[width=\linewidth]{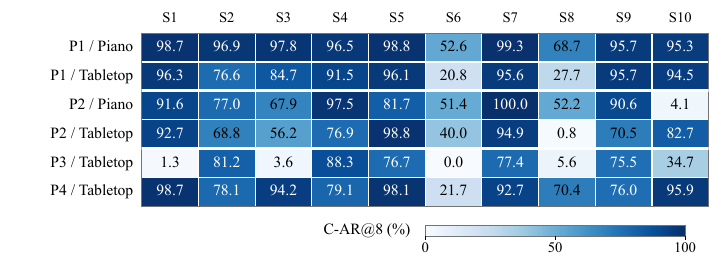}
    \caption{Heat map of C-AR@8 (\%) across four participants (P1 to P4) and excerpts from ten pieces (S1 to S10) on the VSF datasets. P1 and P2 contribute both piano and tabletop performances, while P3 and P4 contribute tabletop performances only, following playback of piano performances by P1 and P2, respectively.}
    \label{fig:participant_piece_heatmap}
\end{figure}

Fig.~\ref{fig:participant_piece_heatmap} shows alignment performance for each participant-piece combination. Some recordings approach 100\%, indicating nearly complete alignment, while others fall below 50\%, reflecting major failures such as losing track partway through a performance. Overall, performance appears to vary more across pieces than across participants or piano/tabletop conditions. However, participant-specific failures also occur; for example, P3's low score on S1 (Bach's \textit{Prelude in C major}, BWV~846) may be related to his distinctive fingering, while omitted tapping gestures during his tabletop performance of S3 (Beethoven's \textit{Piano Sonata No.~8}, III) may contribute to tracking errors.

\begin{figure}[t]
\centering
\includegraphics[width=\linewidth]{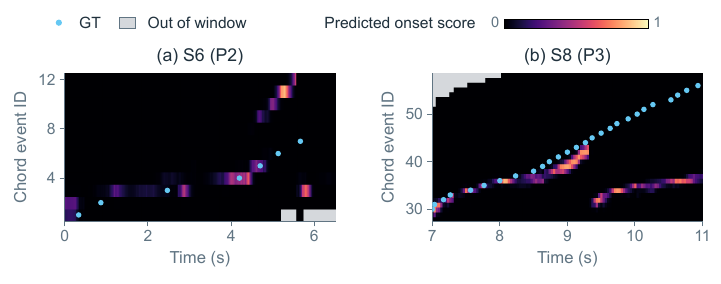}
\caption{Streaming alignment examples on VSF-Tabletop. (a) S6 (P2) exhibits weak or misplaced responses from the beginning. (b) S8 (P3) shows a tracking deviation during a rapid scale passage, where the steeper GT trajectory reflects more closely spaced onsets.}
\label{fig:case_study}
\end{figure}

Among the more challenging pieces, S6 (Debussy's \textit{Clair de lune}) presents difficulties in onset detection from the beginning (Fig.~\ref{fig:case_study}(a)), suggesting that small hand movements during slow passages provide weak visual onset cues. In S8 (Mozart's \textit{Piano Sonata No.~16}, I), tracking failures occur during fast scale passages (Fig.~\ref{fig:case_study}(b)), possibly because the limited number of motion frames per note makes successive movements difficult to distinguish. A higher frame rate may help capture these rapid movements. In contrast, S7 (Petzold's \textit{Minuet in G major}) and S4 (Chopin's \textit{Grande valse brillante}, Op.~18) generally perform better, potentially benefiting from simpler patterns or steady left-hand accompaniment that provides strong alignment cues.

\subsection{Ablation Study}
\label{sec:motion_ablation}

\begin{table}[t]
    \centering
    \caption{C-AR@8 (\%) for ablation variants on the in-domain evaluation set PM60, and test sets VSF-Piano and VSF-Tabletop.}
    \label{tab:ablation}
    \small
    \setlength{\tabcolsep}{4pt}
    \begin{tabular}{clccc}
        \toprule
        Row & Variant & PM60 & VSF-Piano & VSF-Tabletop \\
        \midrule
        1 & \textbf{Ours} & 96.09 & 80.71 & \textbf{67.77} \\
        \midrule
        2 & absolute joint XY & 95.88 & \textbf{82.83} & 58.42 \\
        3 & w/o wrist displacement & 95.14 & 70.20 & 61.49 \\
        4 & w/o hand-window jitter & \textbf{96.63} & 76.44 & 60.03 \\
        \midrule
        5 & w/o IOI feature & 92.79 & 80.29 & 66.17 \\
        6 & w/o IOI scaling & 94.90 & 61.48 & 48.83 \\
        \midrule
        7 & $\rho=0.4$~s, $\varepsilon=0$ & 96.55 & 81.54 & 66.39 \\
        \bottomrule
    \end{tabular}
\end{table}

Table~\ref{tab:ablation} reports ablation results on both VSF datasets, with PM60 included to assess in-domain performance. For motion representation, row~2 replaces the non-wrist XY coordinates with the XY components of $\mathbf{x}_{t,h,j}-\boldsymbol{\mu}_h$ for $j=1,\ldots,20$, where $\boldsymbol{\mu}_h$ is the mean wrist position computed over training data for normalization. Wrist-relative Z is retained because absolute depth estimates can be unstable, while wrist displacement remains unchanged. This improves VSF-Piano performance but reduces VSF-Tabletop performance greatly, supporting the role of wrist-relative coordinates in piano-to-tabletop transfer. In row~3, setting $\Delta\mathbf{x}_{t,h,0}=\mathbf{0}$ reduces overall performance, indicating that wrist translation provides useful information beyond local finger articulation. In row~4, we disable the independent perturbation of motion-segment start and end boundaries by up to 0.5~s. This slightly improves PM60 but lowers performance on both VSF datasets, particularly VSF-Tabletop, suggesting that boundary augmentation may help tolerate inaccurate motion-window placement when alignment predictions are less reliable.

For score representation, disabling IOI scaling augmentation (row~6) substantially reduces VSF performance, while completely disabling the embedding branch for the IOI vector $\mathbf{r}_i$ (row~5) yields similar performance on VSF but lower performance on PM60. This pattern suggests that the model may rely on rhythmic patterns specific to the automatically constructed training scores, reflecting a possible gap between these transcribed scores and real scores used during inference. IOI augmentation may help reduce this gap. Training with performances aligned to real scores is therefore a potential direction for further improvement.

Finally, setting $\rho=0.4$~s and $\varepsilon=0$ (row~7) yields similar performance while preserving the nominal anticipation $\tau=\varepsilon+\rho/2=0.2$~s. Nevertheless, allowing a separate target shift provides flexibility: when the desired anticipation $\tau$ changes, we can jointly adjust $\varepsilon$ and $\rho$ to match it while controlling the temporal spread of supervision.

\subsection{Runtime and Latency}
\label{sec:streaming_performance}

We measure per-frame processing time on a single RTX 4090, using 1,800 frames after a 60-frame warm-up for both WiLoR and AIRPiano inference. WiLoR has mean and 95th-percentile (P95) processing times of 60.45 and 62.13~ms, respectively, compared with 6.68 and 6.88~ms for AIRPiano, which accepts extracted hand poses directly. Hand extraction therefore accounts for most of the processing cost.

At 30~FPS, under ideal parallel execution, distributing alternating frames across two GPUs gives each GPU $2\times1000/30\approx66.7$~ms to process a frame, exceeding WiLoR's P95 extraction time. The P95 processing times measured separately for WiLoR and AIRPiano sum to less than 100~ms. With an anticipation of $\tau=200$~ms, this should leave a nominal budget of over 100~ms for camera capture, audio output, and other overhead. We have implemented a live system and recorded a demonstration using WiLoR's fast inference mode for its greater throughput and latency margins. Comprehensive end-to-end latency measurements and a user study are left to future work.
\section{Conclusion}
\label{sec:conclusion}
We presented AIRPiano, a score-guided system that turns tabletop piano gestures into sound through online motion-score alignment. By combining motion and score representations, supervised alignment learning, and streaming sequence tracking, AIRPiano learns from existing piano performance recordings to support tabletop interaction. Evaluation on VSF-Piano and VSF-Tabletop provides an initial assessment of this approach across piano and tabletop performances.

Our evaluation focuses on alignment accuracy and processing time, which may not fully capture users' experience of playing with live feedback. Future work will therefore refine the live system design and conduct user studies to evaluate perceived latency and compatibility with users' existing playing habits. We will also explore expressive control such as note velocity, note-off timing, ornaments, and onset asynchrony within chords.

\ifanonymous
\else
\section*{Acknowledgments}
We thank the volunteers for contributing piano and tabletop recordings, as well as those who assisted with equipment procurement. We also thank Huiran Yu for recommending relevant literature, and Jiawei Yu and Haina Zhu for their helpful suggestions on manuscript writing.
\fi

\bibliographystyle{splncs04}
\bibliography{references}

\end{document}